\documentclass{elsart}

\usepackage{natbib}
\usepackage{color}

 \usepackage{graphicx}

\usepackage{amssymb}

\begin{document}

\begin{frontmatter}



\title{Heterogeneous distribution of metabolites across plant species}


\author[Tokyo,corr]{Kazuhiro Takemoto}
\ead{takemoto@cb.k.u-tokyo.ac.jp}
\author[Tokyo,PSC,Keio]{Masanori Arita}

\address[Tokyo]{Department of Computational Biology, Graduate School of Frontier Sciences, University of Tokyo, Kashiwanoha 5-1-5, Kashiwa, Chiba 277-8561, Japan}

\address[PSC]{RIKEN Plant Science Center, Suehiro-cho 1-7-22, Tsurumi-ku, Yokohama, Kanagawa 230-0045, Japan}

\address[Keio]{Institute for Advanced Biosciences, Keio University, Baba-cho 14-1, Tsuruoka, Yamagata 997-0035, Japan}

\corauth[corr]{Corresponding author.}

\begin{abstract}
We investigate the distribution of flavonoids, a major category of plant secondary metabolites, across species.
Flavonoids are known to show high species specificity, and were once considered as chemical markers for understanding adaptive evolution and characterization of living organisms.
We investigate the distribution among species using bipartite networks, and find that two heterogeneous distributions are conserved among several families: the power-law distributions of the number of flavonoids in a species and the number of shared species of a particular flavonoid.
In order to explain the possible origin of the heterogeneity, we propose a simple model with, essentially, a single parameter.
As a result, we show that two respective power-law statistics emerge from simple evolutionary mechanisms based on a multiplicative process.
These findings provide insights into the evolution of metabolite diversity and characterization of living organisms that defy genome sequence analysis for different reasons.
\end{abstract}


\end{frontmatter}

\section{Introduction}
Living organisms produce compounds of many types via their metabolisms which are believed to adaptively shape-shift with changing environment across a long evolutionary history.
Elucidation of design principles behind such complex systems is a major goal in natural science.
Toward this end, so far, the structure of metabolic networks has been actively investigated using network analysis from the viewpoint of statistical mechanics.
As a result, striking structural properties such as scale-free (heterogeneous) connectivity and hierarchical organization have been revealed, and possible origins have been discussed via several models (e.g., reviewed in Refs. \cite{Albert2002,Barabasi2004,Albert2005}).
In addition to considering metabolic networks, however, it is also important to consider how metabolites are distributed among species in order to elucidate design principles of metabolisms such as adaptive mechanisms.
The metabolite distributions have the following advantages.
Since living organisms have specific metabolite compositions due to metabolisms adaptively changing with respect to the environment, we can estimate environmental adaptation (adaptive evolution) using metabolite distributions.
Moreover, they are also useful for characterizing species relationships, which are highly linked to ecological systems.
In metabolite distributions, thus, identification of structures and construction of a theory (model) for evolutionary mechanisms are key challenges for a deeper understanding of metabolism.

Flavonoids are especially interesting examples when considering metabolite distributions among species.
Secondary metabolites including flavonoids, alkanoids, terpenoids, phenolics, and other compounds are widely observed in angiosperms, and are not essential for preserving life unlike basic metabolites such as bases, amino acids, sugars, and fatty acids (building blocks of DNA, protein, carbohydrate, and fat, respectively).
However, secondary metabolites play additional roles aiding survival in diverse environments.
Therefore, distributions of secondary metabolites are believed to be significantly different among species due to adaptation to environments, implying high species specificity \cite{Gershenzon1983}.
For this reason, secondary metabolites help us to understand environmental adaptation and adaptive evolution. Moreover, secondary metabolites, especially flavonoids, are often used as markers in chemotaxonomy, which is a taxonomic classification based on metabolite compositions of species that has been used for many years \cite{Bohm1998}.
However, taxonomic classifications using secondary metabolites at higher levels (e.g. family and order levels) are known to be inherently more difficult than those at lower levels (e.g. species levels) \cite{Gershenzon1983}.

Although the metabolite distribution provides important insights into metabolism as discussed above, it has not caught as much attention as metabolic networks.
This was mainly because knowledge of secondary metabolites was not widely available.
In recent years, however, the whole picture of species-flavonoid relationships has become available in the KNApSAcK database \cite{Shinbo2006} and partly in Metabolomics.JP \cite{Arita2008}.
We now can investigate metabolite distributions among species using these websites.

In this paper, we focus on flavonoids, which are a class of secondary metabolites, and investigate metabolite distribution among species.
In order to comprehensibly describe species-flavonoid relationships, bipartite networks are utilized.
They are useful for representing two different objects, which correspond to species and flavonoids in this case.
We first investigate degree distributions in species-flavonoid networks in several families, and show power-law distributions of the number of flavonoids in a species and the number of shared species of a flavonoid.
A simple model is next proposed for explaining a possible origin of the heterogeneous distributions (power-law distributions), and it is compared to real data. 
Furthermore, intuitive descriptions and mathematical evidence are provided for the emergence of heterogeneous distributions.
We finally mention the characteristics of well-shared (hub) isoflavonoids in Fabaceae (bean family) as an example.
In addition, we discuss the possibility of more effective selection of discriminative metabolites and taxonomic classifications at higher levels by considering this heterogeneous distribution, and speculate on the evolution of flavonoid diversity.

\section{Methods}
A total of 14378 species-flavonoid pairs were downloaded from Metabolomics.JP \cite{Arita2008} (http://metabolomics.jp/wiki/Category:FL), in which 4725 species and 6846 identified flavonoid structures are linked by a published journal article.
In other words, only published data were utilized.
The species-flavonoid pairs are by no means comprehensive: no plant species has been `completely' investigated for its biosynthetic activity, and many flavonoid molecules whose descriptions have yet to be published are also thought to exist.

The taxonomy (family) of a species was assigned according to The Taxonomicon (http://taxonomicon.taxonomy.nl).
We here focus on the six largest families in terms of the number of reported flavonoids: Fabaceae (bean family), Asteraceae (composite family), Lamiaceae (Japanese basil family), Rutaceae (citrus family), Moraceae (mulberry family), and Rosaceae (rose family).
In particular, we can discuss in detail species-flavonoid relationships in Fabaceae because this family is well researched.

Flavonoids consist of backbone structures and their modifications.
On the above website, flavonoids are classified into nine groups according to their backbone structures: FL1: Chalcone, FL2: Flavanone, FL3: Flavone, FL4: Dihydroflavonol, FL5: Flavonol, FL6: Flavan, FL7: Anthocyanin, FLI: Isoflavonoid, FLN: Neoflavonoid.

In order to comprehensibly investigate the distribution of flavonoids (metabolites) among species, we here utilize bipartite networks (hereinafter called {\it species-flavonoid networks}), which is useful for representing two different objects (species and flavonoids in this case).
Bipartite networks are defined as graphs having two different node sets (species and flavonoids) in which edges are only drawn between one node set and the other node set (inter-connectivity).
Note that there is no edge between nodes belonging to the same node set (intra-connectivity).
In the species-flavonoid networks, an edge is drawn between a species node and a flavonoid node when the species has the flavonoid.

\section{Results and discussion}
\subsection{Heterogeneous distribution of flavonoids}
\begin{figure}[tbp]
\begin{center}
	\includegraphics{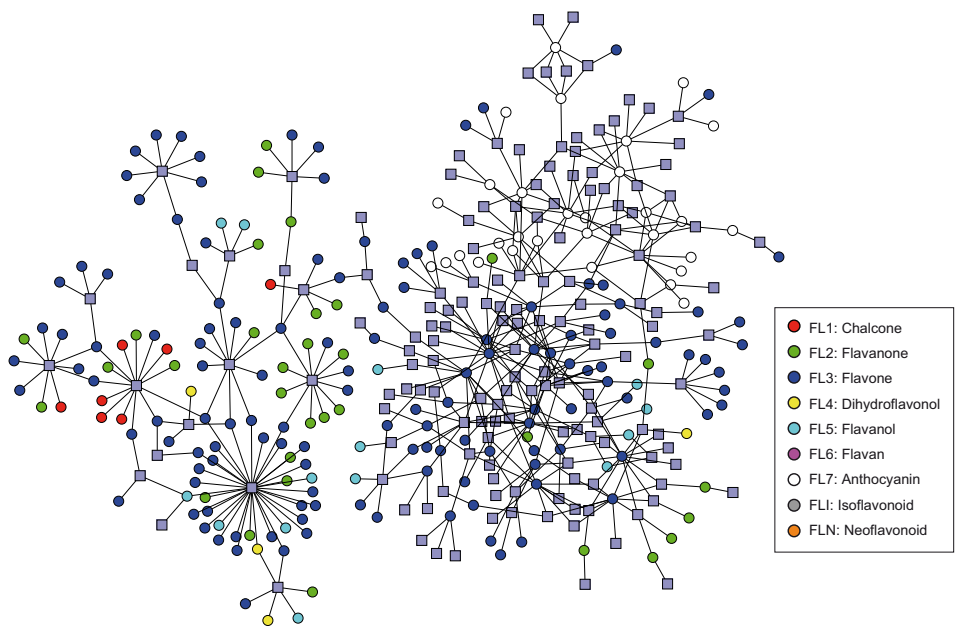}  
	\caption{
	A partial species-flavonoid network for Lamiaceae (the Japanese basil family) drawn by yEd \cite{yed}.
	The squares and circles correspond to plant species and flavonoids, respectively.
	The color of circles indicates the flavonoid class.
	}
	\label{fig:sample_net}
\end{center}
\end{figure}

We show a partial species-flavonoid network for Lamiaceae (the Japanese basil family) in Fig. \ref{fig:sample_net} as an example.
The node degrees of species nodes (squares) and flavonoid nodes (circles) are extremely varied.

In order to characterize the tendency of connectivity, we investigated frequency distributions of the node degree (degree distribution) in species-flavonoid networks.
In bipartite networks, we can find distributions of two types due to the two types of nodes (species nodes and flavonoid nodes).
The number of edges for species nodes and flavonoid nodes correspond to the number of flavonoids in a species $n_f$ and the number of shared species of a particular flavonoid $n_s$, respectively.

\begin{figure}[tbp]
\begin{center}
	\includegraphics{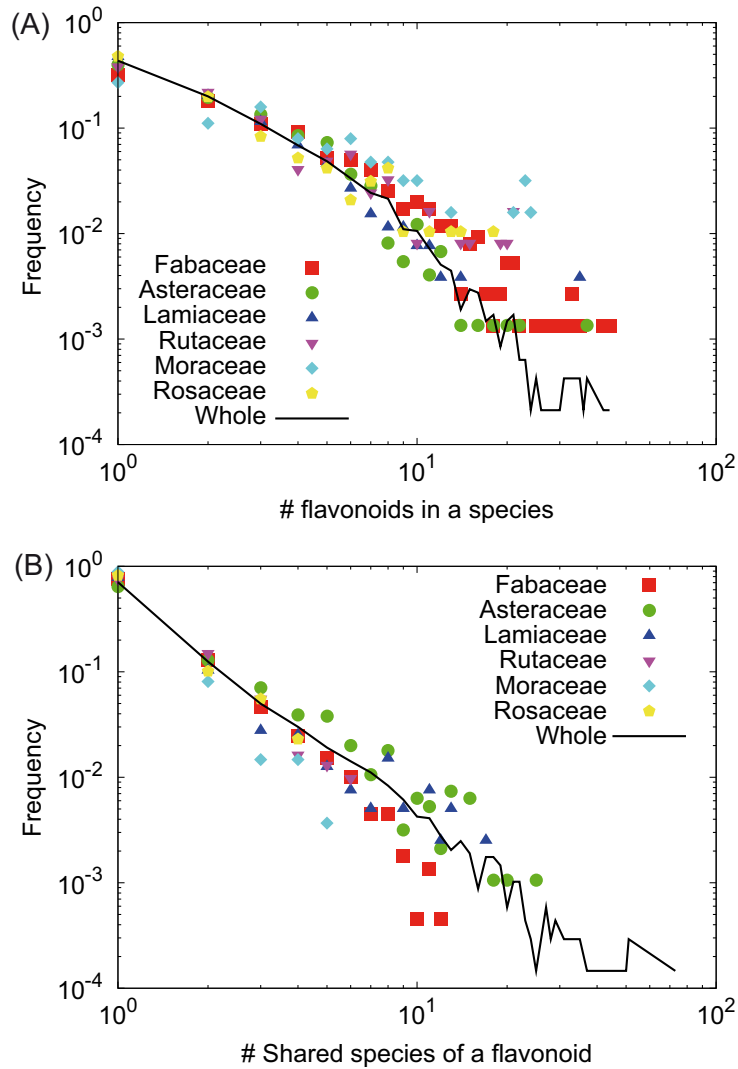}  
	\caption{
	Degree distributions of the species-flavonoid networks.
	(A) The frequency distribution of the number of flavonoids in a plant species.
	(B) The frequency distribution of the number of shared species of a flavonoid.
	The symbols indicate the distributions for family-based species-flavonoid networks.
	The solid lines represent the distributions for all-encompassing species-flavonoid networks.
	}
	\label{fig:degree_dist}
\end{center}
\end{figure}

As shown in Fig. \ref{fig:degree_dist}, the frequency distributions of $n_f$ and $n_s$ roughly follow a power law, implying heterogeneous distribution of flavonoids among species.
That is, most flavonoids are shared by a few species; however, a few flavonoids are conserved in many species.
Most species have flavonoids of a few types; however, a few species have flavonoids of many types.
Furthermore, the heterogeneous distributions of flavonoids among species characterized by the power-law statistics are approximately conserved between family-based species-flavonoid networks, suggesting a scale-free feature.
Regarding the number of metabolites in a species following power-law distributions a similar result has been additionally reported in \cite{Shinbo2006}.

This finding might provide insights into environmental adaptation and adaptive evolution because compositions of secondary metabolites including flavonoids are strongly influenced by environmental conditions.
By considering heterogeneous distribution of flavonoids, we might be able to detect useful metabolites relating to such adaptations.
In particular, the heterogeneous distribution of the number of shared species predicts that most flavonoids are important for characterizing such adaptations at a species level.

However, this finding also tells us of the difficulty in finding metabolites that reflect adaptations at higher levels (e.g. family and order levels).
In general, taxonomic classification (characterization of living organisms) based on secondary metabolites including flavonoids is empirically believed to be more difficult at higher levels than at lower levels (e.g. species level) \cite{Gershenzon1983}.
This difficulty, driving from the heterogeneous distributions is that most flavonoids are species-specific.
Thus, most flavonoids contribute to taxonomic classification at lower levels rather than that at higher levels.
As a result, we possibly extract overrated and/or underrated information when exhaustively considering flavonoids (metabolites) at higher levels.
Hub flavonoids might play an essential role in characterizing biological features (e.g. environmental adaptation) at higher levels because of their conservation; thus, hub flavonoids are expected to be more appropriate and beneficial markers for characterization of biological features at higher levels.

\subsection{A possible origin of heterogeneous distributions}
We here speculate on a possible origin of heterogeneous distributions of flavonoids (metabolites) among species using a simple model.
We believe that the heterogeneous distribution was acquired through evolutionary history.
The origin of heterogeneity might provide insights into the evolution of flavonoid diversity.

\subsubsection{Model}
We consider two simple evolutionary mechanisms as follows.
(i) New flavonoids are generated by variation of existing flavonoids.
In evolutionary history, species accordingly obtain new metabolic enzymes via gene duplications \cite{Diaz-Mejia2007} and horizontal gene transfers \cite{Pal2005}, and the metabolic enzymes synthesize new flavonoids through modification of existing flavonoids with substituent groups and functional groups.
(ii) Flavonoid compositions of new species are inherited from those of existing (ancestral) species.
New species are believed to emerge by mutation of ancestral species, and they are similar to the ancestral species as a result.
For this reason, there might be the above inheritance of flavonoid compositions from ancestral species to new species.

With consideration for the above two mechanisms, we propose a simple model with two parameters $p$ and $q$ reproducing heterogeneous distributions of flavonoids among species.

\begin{figure}[tbp]
\begin{center}
	\includegraphics{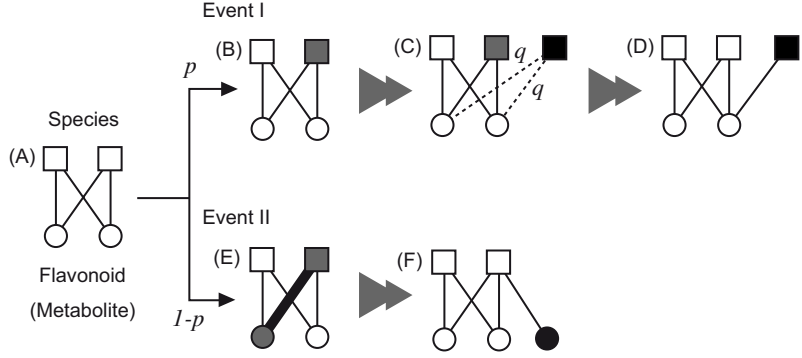}  
	\caption{
	Schematic diagram of the model.
	Squares and circles mean plant species and flavonoids, respectively.
	(A) An initial species-flavonoid relationship (network) with $n_0=2$.
	((B) -- (D)) Event I: the emergence of a new species.
	The gray square represents a randomly selected existing species.
	The black square indicates a new species emerging due to duplication of existing species.
	The dashed lines are possible pairs of the new species and flavonoids.
	((E) -- (F)) Event II: the emergence of a new flavonoid.
	The thick edge between gray nodes corresponds to a randomly selected existing species-flavonoid pair.
	The black circle is a new flavonoid.
	}
	\label{fig:model}
\end{center}
\end{figure}

Our model is defined by the following procedure:

(a) We set an initial species-flavonoid network represented as a complete bipartite graph with $n_0$ species and $n_0$ flavonoids (Fig. \ref{fig:model} A).

(b) With the probability $p$, Event I corresponding to the emergence of a new species occurs.
An existing species is selected at random (Fig. \ref{fig:model} B).
A new species emerges due to mutation of the randomly selected existing species, and the flavonoids of the existing species are inherited by the new species as their candidate flavonoids (Fig. \ref{fig:model} C).
On considering divergence of flavonoid compositions, the new species finally acquires flavonoids with equal probability $q$ for each of the candidates (Fig. \ref{fig:model} D).
However, if new species accordingly have no flavonoids, then such species are neglected (removed) because of the observation condition (species without flavonoids are not included in our data set).
In contrast to Event I, Event II corresponding to the emergence of a new flavonoid occurs with the probability $1-p$.
A species-flavonoid pair is uniformly selected at random (Fig. \ref{fig:model} E). 
Then, the species receives a new flavonoid (Fig. \ref{fig:model} F).

(c) The procedure (b) is repeated until the number of species and the number of flavonoids are equivalent to $S$ and $F$, respectively.

This model does not consider the loss of nodes
(extinctions), which is an important mechanism in evolutionary systems.
In particular, the degree distributions may become different due to such extinctions \cite{Enemark2007,Deng2007}.
In species-flavonoid networks, however, this mechanism tends to be nonessential for the following reason.
In plant evolution, genome doubling (polyploidity) is a major driving force for increasing genome size and the number of genes \cite{Adams2005}.
Duplicated genes typically diversify in their function, and some acquire the ability to synthesize new compounds.
Indeed, more than 50,000 molecular structures are elucidated in the entire plant kingdom (mostly secondary metabolites) \cite{DeLuca2000}, compared to a few thousand primary metabolites in higher animals.
The population of flavonoids, the representative group in plant secondary metabolites, is therefore expected to increase, indicating that we can dismiss the effect of node losses.

Our model is essentially adjusted through only one parameter $q$ because the parameter $p$ only controls the number of species and the number of flavonoids.
When we set the number of species $S$ and the number of flavonoids (metabolites) $F$, the parameter $p$ can be estimated as $S/(S+F)$.

\subsubsection{Relation with `rich-get-richer' mechanisms}
The emergence of heterogeneous (power-law) distributions in evolving systems might be caused by `rich-get-richer' or preferential mechanisms: the increase of a statistic is proportional to the statistic itself \cite{Barabasi2004,Barabasi1999}.
We here explain that the model has `rich-get-richer' mechanisms.

We first mention the number of flavonoids in a species $n_f$.
When Event II occurs, $n_f$ increases.
The number of flavonoids of species $i$, $n_f^i$, increases when a randomly selected species-flavonoid pair includes species $i$.
Thus, species with many flavonoids tend to be selected in such a case.
As a result, such species acquire more flavonoids, implying a `rich-get-richer' mechanism.
The origin of this preferential mechanism is similar to that in the Dorogovtsev-Mendes-Samukhin (DMS) model \cite{Dorogovtsev2001}.
However, our model is essentially different from the DMS model because the DMS model does not describe bipartite relationships.

This is mathematically described as follows.
We consider the time evolution of $n_f^i$. Let $L(t)$ be the total number of species-flavonoid pairs at time $t$; the probability that species $i$ with $n_f^i$ flavonoids is chosen is equivalent to $n_f^i/L(t)$ because the pair is randomly selected.
In addition, Event II occurs with the probability $1-p$.
Therefore, the time evolution of $n_f^i$ is described as
\begin{equation}
\frac{d}{dt}n_f^i=(1-p)\frac{n_f^i}{L(t)}.
\label{eq:dn_f/dt}
\end{equation}

Moreover, we focus on the time evolution of $L(t)$.
The number of pairs $L(t)$ increases in Events I and II.
In the case of Event I, $L(t)$ increases by $q\times L(t)/S(t)$, where $S(t)$ is the number of species at time $t$, because flavonoids of the randomly selected existing species are inherited by the new species with the probability $q$.
Note that the expected number of flavonoids of randomly selected species is $\sum_{j=1}^{S(t)}n_f^j/S(t)=L(t)/S(t)$.
In the case of Event II, $L(t)$ increases by 1.
The events I and II occur with the probabilities $p$ and $1-p$, respectively.
Therefore, the time evolution of $L(t)$ is written as
\begin{equation}
\frac{d}{dt}L(t)=pq\frac{L(t)}{S(t)}+(1-p).
\end{equation}
Since $S(t) = pt$, the solution of this equation with the initial condition $L(1)=L_0$ is
\begin{eqnarray}
L(t) &=&\frac{1-p}{1-q}t+\left(L_0-\frac{1-p}{1-q}\right)t^q \\
&\approx& \frac{1-p}{1-q}t \ \ \ (t\gg 1),
\label{eq:approx_L}
\end{eqnarray}
indicating that $L(t)$ is approximately proportional to time $t$ for large $t$ and relatively small $q$.

Substituting Eq. (\ref{eq:approx_L}) into Eq. (\ref{eq:dn_f/dt}), we have
\begin{equation}
\frac{d}{dt}n_f^i\approx(1-q)\frac{n_f^i}{t},
\label{eq:dn_f/dt2}
\end{equation}
suggesting the preferential mechanism: the increase of $n_f^i$ is proportional to $n_f^i$.

From this equation, using the mean-field-based method \cite{Barabasi1999}, we immediately obtain the power-law distribution of $n_f$:
\begin{equation}
P(n_f)\sim n_f^{-(2-q)/(1-q)}.
\end{equation}

We next consider the number of shared species of a flavonoid $n_s$.
When Event I occurs, $n_s$ increases.
The number of shared species of a flavonoid $i$, $n_s^i$, might increase when a randomly selected species has flavonoid $i$.
Thus, flavonoids shared by many species tend to be selected in such a case.
As a result, such flavonoids are shared by more species, reflecting a `rich-get-richer' mechanism.
The origin of this preferential mechanism is analogous to that in the duplication-divergence (DD) model \cite{Vazquez2003-1,Vazquez2003-2}.
However, our model is also different from the DD model in that the DD model does not describe bipartite relationships.

This is mathematically explained as follows.
We consider the time evolution of $n_s^i$.
The probability that each flavonoid $i$ is shared by a new species is equivalent because a resulting new species obtains flavonoid $i$ with the probability $q$ when one of $n_s^i$ species with flavonoid $i$ is randomly selected.
In addition, Event I occurs with the probability $p$. Since $S(t)=pt$, therefore, the time evolution of $n_s^i$ is described as
\begin{equation}
\frac{d}{dt}n_s^i=pq\frac{n_s^i}{S(t)}=q\frac{n_s^i}{t},
\label{eq:dn_s/dt}
\end{equation}
implying the preferential mechanism: the increase of $n_s^i$ is proportional to $n_s^i$.

Like for $n_f$, we immediately have the power-law distribution of $n_s$:
\begin{equation}
P(n_s)\sim n_s^{-(1+q)/q}.
\end{equation}

\subsubsection{Comparison with real data}
We compare frequency distributions of the number of flavonoids of a plant species $n_f$ and the number of shared species of a flavonoid $n_s$ between our model and real data.
We consider the frequency distributions obtained from the whole data set.

The parameter $p$ is derived from $S/(S+F)$. Since $S=4725$ and $F=6846$ in the actual species-flavonoid network, the parameter $p$ becomes 0.41.

\begin{figure}[tbp]
\begin{center}
	\includegraphics{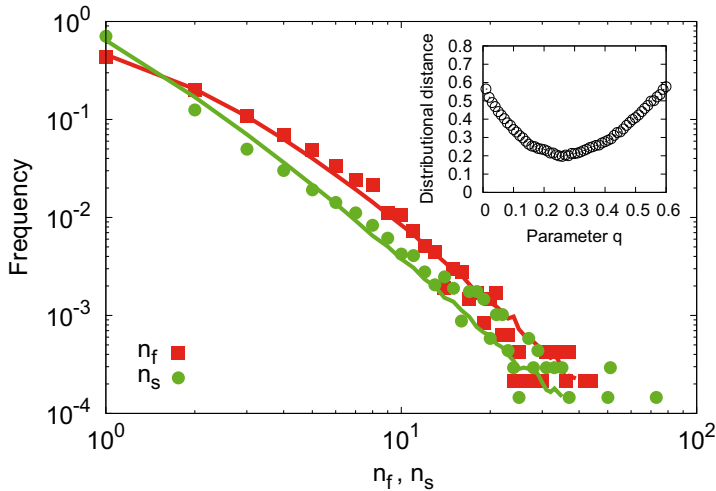}  
	\caption{
	Comparison of frequency distributions between real data and the model.
	The red squares represent the real frequency distribution of the number of flavonoids in a plant species $n_f$.
	The green circles indicate the real frequency distribution of the number of shared species of a flavonoid $n_s$.
	The solid lines correspond to the predicted frequency distributions averaged over 100 realizations of the model with $q=0.26$.
	The inset shows the distributional distance between predicted distributions and empirical distributions with the parameter $q$.
	}
	\label{fig:comparison}
\end{center}
\end{figure}

\begin{figure}[tbp]
\begin{center}
	\includegraphics{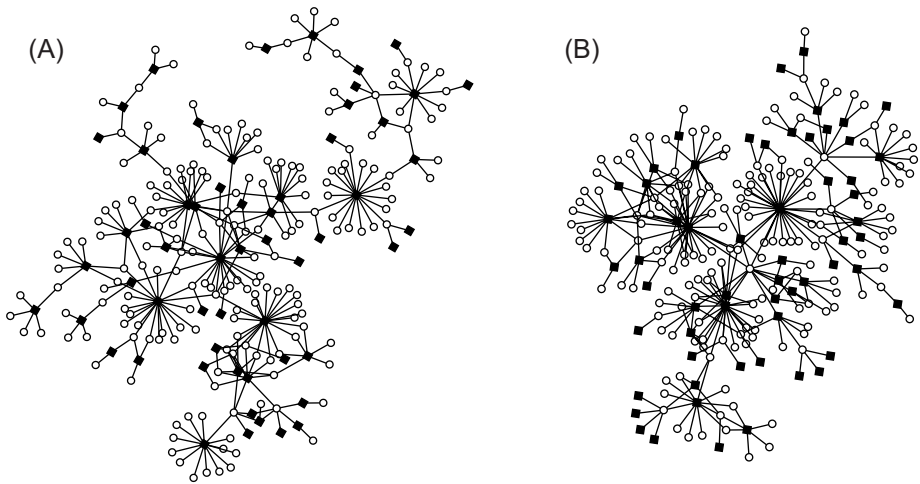}  
	\caption{
	Comparison of network structure between real data for Rutaceae (A) and our model with $q=0.26$ (B).
	Filled squares and open circles correspond to plant species and flavonoids, respectively.
	The networks are drawn by yEd \cite{yed}.
	}
	\label{fig:net_compari}
\end{center}
\end{figure}

Fig. \ref{fig:comparison} shows the comparison of the frequency distribution between real data and the model with $q=0.26$.
The parameter $q$ is selected by minimizing the distributional distance (the inset in Fig. \ref{fig:comparison}), which corresponds to the sum of tail-weighted Kolmogorov-Smirnov statistics (distances) \cite{Press1992} for two distributions [$P(n_f)$ and $P(n_s)$] between the predicted distributions and the empirical distributions.
Fig. \ref{fig:net_compari} shows the comparison of the network structure between real data and our model.
Here, we chose the species-flavonoid network for Rutaceae (the citrus family) as an example because of its reasonable size.

As shown in these figures, the model is in good agreement with real data, indicating that the two mechanisms in our model are a possible origin of the heterogeneous distribution.
The power-law statistics are conserved among families as shown in Fig. \ref{fig:degree_dist}, suggesting that the evolutionary mechanisms might be universal among families.

In our model, the parameter $q$ means the probability that flavonoid compositions of new species, which emerge due to mutations of an existing (ancestral) species, are inherited from ancestral species.
As above, the inheritance probability is relatively small ($q=0.26$), suggesting high divergence of flavonoid composition from ancestral species.
This might be because new species eventually acquire adaptive compositions of flavonoids different from those ancestral species in different environments in evolutionary history.
In particular, species' habitats might be strongly reflected in the composition of secondary metabolites because such metabolites play crucial roles in survival in diverse environments.
For this reason, flavonoid compositions are expected differ between new species and ancestral species.
In this manner, our model can estimate the divergence degree of flavonoids from metabolite distributions.

In addition, our model might also be applied to wide-range metabolites among species because the emergence mechanisms, considered in the model, are possibly also similar in the case of other metabolites such as other secondary metabolites (e.g. alkanoids, terpenoids, and phenolics) and lipids, which are important from the viewpoints of pharmacodynamics and dietetics.
Thus, we can predict that heterogeneous distributions are observed in not only flavonoids but also other metabolites.
Furthermore, our model also helps to estimate the divergence degree of such metabolites.

\subsection{Characteristics of hubs in species-flavonoid networks: the case of Fabaceae}
We here investigate hubs in species-flavonoid networks.
In particular, hubs for flavonoid nodes might be key characteristics at higher levels because such flavonoids are well conserved among species at higher levels.
As an example, we focus on Fabaceae (the bean family), already well investigated as the source of isoflavonoids.

Hubs are defined using a $Z$-score that characterizes disagreement with an average \cite{Guimera2005,Guimera2007}.
If network connectivity is determined at random like in a random graph, the degree distribution approximately follows a normal distribution.
Then, we can find nodes with large numbers of edges, which are hardly in agreement with the normal distribution, using the $Z$-score.
Thus, we define hubs as nodes with more than $\bar{k}+\sigma z_c$ edges, where $\bar{k}$ is the average number of edges over all flavonoid nodes or species nodes and $\sigma$ is the standard deviation of $\bar{k}$.
$z_c$ is a threshold value used to determine hubs, and we set $z_c=2.5$ \cite{Guimera2005,Guimera2007}.

\begin{figure}[tbp]
\begin{center}
	\includegraphics{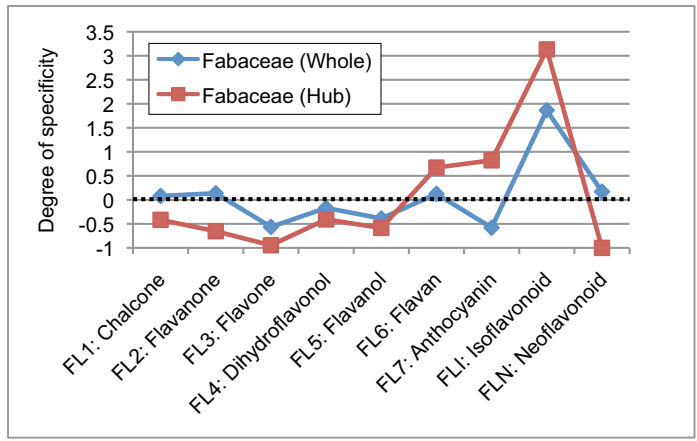}  
	\caption{
	Degree of specificity of each flavonoid class.
	The case of all flavonoids [the item ``Fabaceae (whole)"] and the case of hub flavonoids [the item ``Fabaceae (Hub)"] are shown.
	}
	\label{fig:specificity}
\end{center}
\end{figure}

Fig. \ref{fig:specificity} shows the degree of specificity $S_i$ for each flavonoid class, defined as $S_i=r_i/R_i-1$ where $r_i$ is the ratio of flavonoid class $i$ in a target flavonoid set, and $R_i$ is the ratio of flavonoid class $i$ in the whole data set.
A positive $S_i$ indicates a discriminative flavonoid class $i$.

As shown in this figure, there are significantly many isoflavonoids.
It is well known that isoflavonoids are predominantly found in Fabaceae \cite{Veitch2007}, in good agreement with the result obtained from our analysis.
The presence of dominant flavonoids is also predicted by our model.
Species have many similar flavonoids due to the evolutionary mechanism: new flavonoids are generated by modification of existing (ancestral) flavonoids.
This result also supports the reliability of our model.

In particular, the significance of isoflavonoids is clearer in the case of hub flavonoids.
Here, we can also observe discriminative anthocyanins.
This class of flavonoid serves as pigments in plants, and some are well conserved across many species.
In Fabaceae, thus, we also expect to characteristically observe anthocyanins as typified by chrysanthemin in black beans.
In the case of whole flavonoids, however, anthocyanins are relatively less conspicuous, unlike the case for hub flavonoids.

As explained in the previous section, we have possibilities of extracting overrated and/or underrated information using whole flavonoids due to the heterogeneous distribution.
To detect higher-level-specific (e.g. family-specific) characters based on whole flavonoids, we need to assume the number of shared species of a flavonoid with a normal (homogeneous) distribution.
However, the statistics follow a power-law distribution as shown in Fig. \ref{fig:degree_dist} B, indicating that the assumption is not appropriate.
Therefore, we might extract more appropriate information at higher levels from hub (well-shared) flavonoids rather than whole flavonoids.

As an example, let us show hubs of isoflavonoids significantly more distributed within Fabaceae. Table \ref{table:hub_flavonoid} shows the list of the top 20 isoflavonoids in Fabaceae ranked by the number of shared species.
As shown in this table, the hub flavonoids are shared among species of diverse types.
In hub flavonoids, moreover, we find genistein and daizein, which are believed to be origins of all isoflavones \cite{Veitch2007}.
Most of the other hub flavonoids are synthesized via simple modifications of genistein and daizein.
This result might support a hypothesis from our model: hub (well-shared) flavonoids are ancient.

\begin{table}[tbp]
\caption{List of the top 20 isoflavonoids in Fabaceae ranked by the number of shared species.}
\begin{center}
\includegraphics{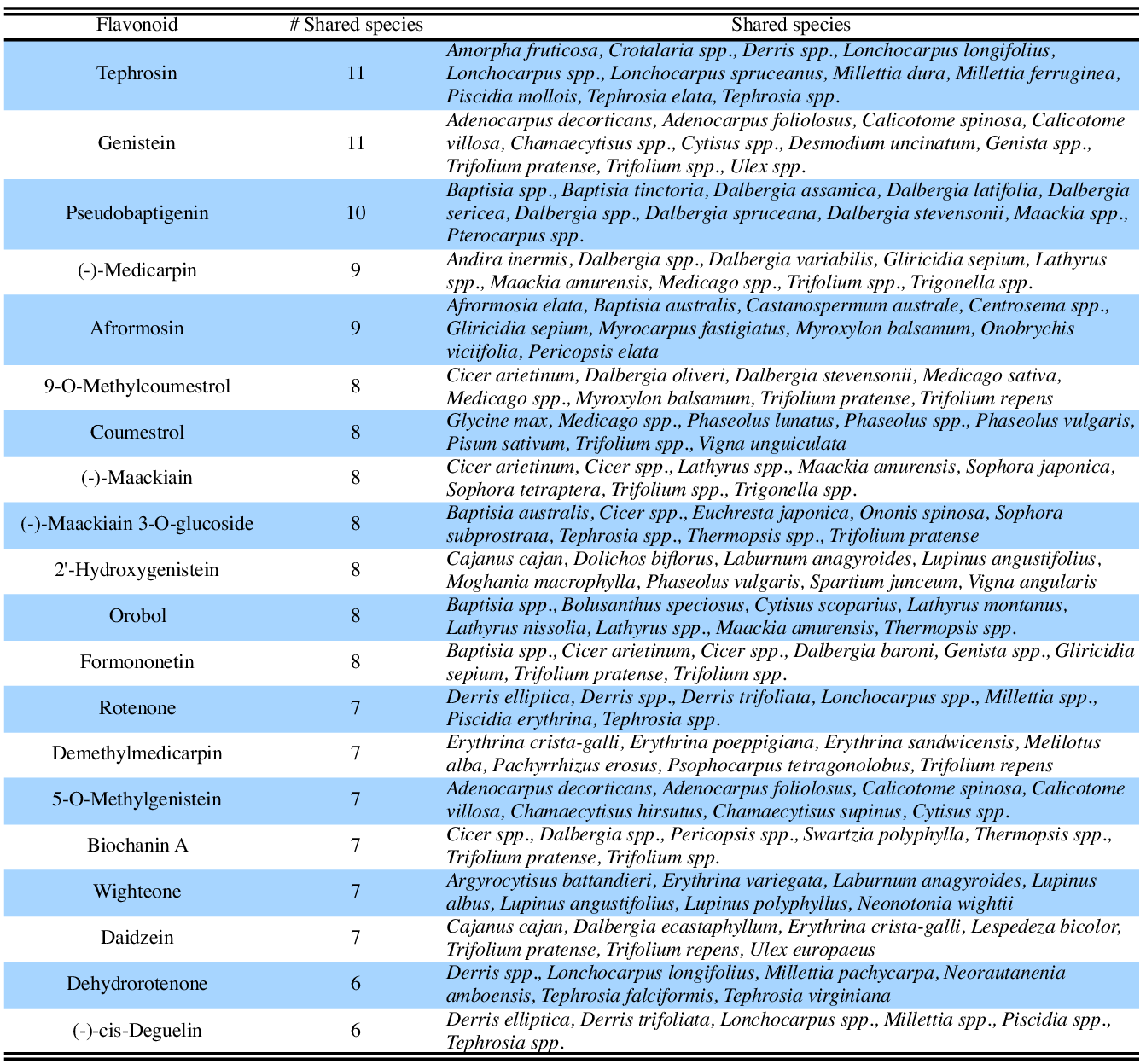}
\end{center}
\label{table:hub_flavonoid}
\end{table}

\begin{figure}[tbp]
\begin{center}
	\includegraphics{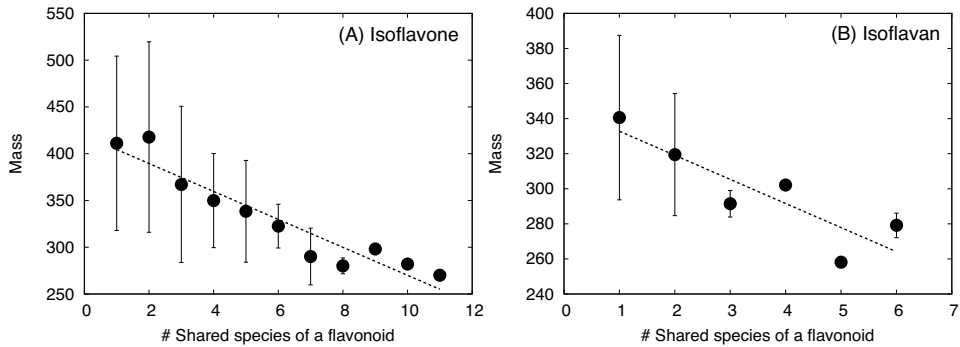}  
	\caption{
	Correlations between masses of flavonoids and the number of shared species $n_s$.
	(A) Isoflavone (Spearman's rank correlation $r=-0.23$ with $P<10^{-5}$).
	(B) Isoflavan (Spearman's rank correlation $r=-0.33$ with $P<0.01$)
	}
	\label{fig:mass}
\end{center}
\end{figure}

Our model has an assumption: new flavonoids are generated by modification of existing (ancestral) flavonoids.
If this assumption is appropriate, we can expect hub (well-shared) flavonoids  to have relatively simple modifications.
As explained in the previous section, flavonoids consist of backbone structures and their modifications.
Thus, we can estimate the complexity of modification for flavonoids using their masses in the case where two flavonoids share backbone structures.
In order to test this hypothesis, we investigate correlations between masses of flavonoids and the number of shared species $n_s$ (Fig. \ref{fig:mass}).
As an example, we here focus on backbone structures of two types for isoflavonoids: isoflavone and isoflavan (highly observed in Fabaceae).
As shown in this figure, there are negative correlations, implying relatively simple modification of hub flavonoids.
This result suggests validity of the assumption in the model.
In addition, this result also suggests that we can predict the number of shared species of a flavonoid using its mass.
This might be useful for detecting well-shared flavonoids.

\section{Conclusion}
We have found heterogeneous distributions of flavonoids among species, and this is conserved among several families: fat-tailed distributions of the number of flavonoids in a species and the number of shared species of a flavonoid.
In particular, we can extract more appropriate and beneficial flavonoid compositions with consideration for the heterogeneous distribution.
This finding might be useful for taxonomic classification and characterization of living organisms using secondary metabolites including flavonoids at higher levels.
Furthermore, a simple model has been proposed for describing a possible origin of the heterogeneous distribution.
It has been shown that the `rich-get-richer' mechanisms inducing heterogeneous distributions are led by simple evolutionary mechanisms.
The model estimates the divergence degree of several metabolites including flavonoids, and it predicts heterogeneous distribution of such metabolites among species.
We furthermore have found relatively simple modifications of well-shared flavonoids via the model.
Our model helps with understanding the evolution of metabolite diversity.

\section*{Acknowledgment}
We thank J.B. Brown who kindly helped with the proofreading of this paper.
K.T. was partially supported by a Research Fellowship for Young Scientists from the Japan Society for the Promotion of Science.
The work was also supported by a Grant-in-Aid for Scientific Research on Priority Area ``Systems Genomics" from the Ministry of Education, Culture, Sports, Science and Technology, Japan.

 

\end{document}